\documentclass[%
 reprint,
 amsmath,
 amssymb,
 aps,
pra,
]{revtex4-1}

\usepackage{graphicx}
\usepackage{dcolumn}
\usepackage{bm}

\begin{document}

\title{Boosting spatial resolution by incorporating periodic boundary conditions into single-distance
hard-x-ray phase retrieval
}

\author{David M. Paganin}
\affiliation{School of Physics and Astronomy, Monash University, Victoria 3800, Australia}

\author{Vincent Favre-Nicolin, Alessandro Mirone, Alexander Rack, Julie Villanova}
\affiliation{The European Synchrotron -- ESRF, CS40220, 38043 Grenoble, France}

\author{Margie P. Olbinado}
\affiliation{Paul Scherrer Institut,
Forschungsstrasse 111,
5232 Villigen PSI,
Switzerland}

\author{Vincent Fernandez}
\affiliation{The Natural History Museum,
Cromwell Road, London SW7 5BD, United Kingdom}

\author{Julio C. da Silva}
\affiliation{The European Synchrotron -- ESRF, CS40220, 38043 Grenoble, France and NEEL Institute CNRS/UGA, 38042 Grenoble, France}

\author{Daniele Pelliccia}
\affiliation{Instruments \& Data Tools Pty Ltd, Victoria 3178, Australia}

\date{\today}

\begin{abstract}
A simple coherent-imaging method due to Paganin {\em et al.}~is widely employed for phase--amplitude reconstruction of samples using a single paraxial x-ray propagation-based phase-contrast image.  The method assumes that the sample-to-detector distance is sufficiently small for the associated Fresnel number to be large compared to unity. The algorithm is particularly effective when employed in a tomographic setting, using a single propagation-based phase-contrast image for each projection.  Here we develop a simple extension of the method, which improves the reconstructed contrast of very fine sample features.  This provides first-principles motivation for boosting fine spatial detail associated with high Fourier frequencies, relative to the original method, and was inspired by several recent works employing empirically-obtained Fourier filters to a similar end. 
\end{abstract}

\maketitle

\section{Introduction}

In 2002 a simple algorithm was published for reconstructing the projected thickness of a single-material sample given a single propagation-based phase contrast image obtained in the small-defocus regime \cite{Paganin2002}. In this method, the ratio of the real part of the projected refractive index decrement and the projected linear attenuation coefficient is assumed to be both known and constant.  The method assumes paraxial coherent radiation or matter waves (e.g.~x-rays, visible light, electrons or neutrons), plane-wave illumination of known intensity, and an object-to-detector propagation distance that is sufficiently small for each structure in the sample to produce no more than one Fresnel-diffraction fringe \cite{Wilkins1996} (more precisely, the object-to-detector distance is assumed to be small enough to make the corresponding  Fresnel number \cite{SalehTeichBook} large compared to unity).  Within its domain of validity (single-material sample and small object-to-detector propagation distance for paraxial radiation or matter waves), the method may be viewed as providing a computationally-simple unique closed-form deterministic solution to the twin-image problem of inline holography \cite{Gabor1948}, since propagation-based phase contrast images are synonymous with inline holograms \cite{Pogany1997}. 

The 2002 algorithm has been widely utilised, particularly for propagation-based x-ray phase contrast imaging. Its advantages, bought at the price of the previously stated strong assumptions, include simplicity, speed, significant noise robustness even for strongly absorbing samples, and the ability to process time-dependent images frame-by-frame.  Efficient computer implementations are available in the following software packages: ANKAphase \cite{Weitkamp2011}, X-TRACT \cite{Gureyev2011}, pyNX \cite{FavreNicolin2011}, PITRE \cite{Chen2012}, Octopus \cite{Boone2012,Dierick2004}, pyHST2 \cite{mirone2014}, TomoPy \cite{Gursoy2014,Pelt2016}, SYRMEP Tomo Project \cite{Brun2017} and HoloTomo Toolbox \cite{Lohse2020}. While most applications to date have employed x-rays, the method was originally developed with a broader domain of applicability in mind, including but not limited to electrons, visible light and neutrons \cite{Paganin2002}.  Accordingly, the method has now been applied to out-of-focus contrast images \cite{CowleyBook} obtained using electrons \cite{Liu2011}, visible light \cite{Poola2017} and neutrons \cite{Paganin2019}.

When the method of \citet{Paganin2002} (PM)  is utilised in a tomographic context \cite{Mayo2003}, its domain of utility broadens since many objects may be viewed as locally composed of a single material of interest, in three spatial dimensions, even though they cannot be described as composed of a single material in projection \cite{Beltran2010,Beltran2011}.  Examples of applications of the PM in a tomographic setting include the imaging of paper \cite{Mayo2003}, polymer micro-wire composites \cite{Mayo2006}, high-Weber-number water jets \cite{Wang2006}, self healing thermoplastics \cite{Mookhoek2010}, paint-primer micro-structure \cite{Yang2010}, 
sandstone micro-structure \cite{Yang2013}, granite \cite{Denecke2011}, melting snow \cite{Uesugi2012}, anthracite coal \cite{Wang2013},  evolving liquid foams \cite{Mokso2013}, iron oxide particles in mouse brains \cite{Marinescu2013, Rositi2013}, rat brains \cite{Beltran2011}, mouse lungs \cite{Lovric2013}, rabbit lungs \cite{Beltran2011},  mouse tibiae \cite{Stevenson2010}, crocodile teeth \cite{Enax2013}, mosquitoes \cite{Weitkamp2011}, fly legs \cite{Mayo2003}, high speed {\em in vivo} imaging of a fly's flight motor system \cite{Schwyn2013}, wood \cite{Mayo2006}, dynamic crack propagation in heat treated hardwood \cite{Gilani2013}, rose peduncles \cite{Matsushima2012}, amber-fossilised spiders \cite{McNeil2010,Penney2012}, amber-fossilised centipedes \cite{Edgecombe2012}, fossilised rodent teeth \cite{Rodrigues2012}, fossil bones \cite{Sanchez2012}, ancient cockroach coprolites \cite{Vrskansky2013}, fossilised early-animal embryos \cite{Yin2013}, fossil muscles of primitive vertebrates \cite{Trinajstic2013,Sanchez2013} and the vertebral architecture of ancient tetrapods \cite{Pierce2013}.  The preceding list is restricted to papers published prior to 2014.  From 2014 onwards,  several hundred papers have employed the PM for phase-contrast x-ray tomography \footnote{For a partial list of additional references that use the ANKAPhase implementation \cite{Weitkamp2011} of the phase-retrieval algorithm of \citet{Paganin2002}, see e.g.~\texttt{http://www.alexanderrack.eu/ANKAphase/ ankaphase$\textunderscore$users.html}.}.

The present work was inspired by several publications that incorporate unsharp masking \cite{AdrianSheppard2004} or related techniques to boost fine spatial detail in reconstructions obtained using the PM.  These include the deconvolution filter in the ANKAphase \cite{Weitkamp2011} version 2.1 implementation of the PM, incorporation of an unsharp mask into the pyHST2 implementation of the PM \cite{Sanchez2012,mirone2014}, and utilisation of the measured phase contrast image as a physical unsharp mask \cite{Irvine2014}.  These extensions of the method all suppress high spatial-frequency information by a factor less than that given by the Fourier-space Lorentzian \footnote{We use the term ``Lorentzian'' to refer to functions of the form $f(x,a)=1/(a^2 + x^2)$, where $a$ is a real non-zero constant and $x$ is real variable.  Note, however, that such functions are also often referred to as Breit--Wigner distributions or  Cauchy distributions.} filter that is employed in the PM.  Notable also is the work of \citet{Yu2017}, which enhances fine spatial detail by adapting the PM to a multi-image setting. The resulting improvements, most particularly in fine spatial detail obtained via tomographic reconstructions utilising the method, are clearly evident in the previously cited publications.  These publications \cite{Weitkamp2011,Sanchez2013,mirone2014,Irvine2014,Yu2017} provide impetus to revisit the theory underpinning the PM, thereby seeking a first-principles justification for reduced suppression of high spatial frequency information, relative to the Fourier filter in the original form of the method. 

The remainder of the paper is structured as follows.  Section II derives a generalised form of the PM (``GPM''), showing how it reduces to the original form of the single-image phase-retrieval algorithm for low spatial frequencies in the input phase-contrast image data.  Simulated x-ray data is considered in Sec.~III, comparing the GPM to the PM.  Section IV presents two experimental demonstrations of the method.  Section V discusses the domain of applicability for both the GPM and the PM, together with the effective high-pass filter to the PM that is implied by the GPM.  Section VI indicates some possible avenues for future work.  We conclude with a brief summary in Sec.~VII.

\section{Incorporation of periodic boundary conditions into the single-distance phase retrieval}

For a monochromatic scalar x-ray wave-field with intensity $I(x,y,z)$ and phase $\varphi(x,y,z)$ that is paraxial with respect to an optical axis $z$, the associated continuity equation is known as the transport-of-intensity equation \cite{Teague1983} (TIE): 
\begin{eqnarray}
\nabla_{\perp}\cdot[I(x,y,z)\nabla_{\perp}\varphi(x,y,z)]=-k\frac{\partial I(x,y,z)}{\partial z}.
\label{eq:A}
\end{eqnarray}
\noindent Here, $(x,y)$ denote Cartesian coordinates in planes perpendicular to the optical axis, $\nabla_{\perp}$ denotes the gradient operator in the $xy$ plane and $k=2\pi/\lambda$ is the wave-number corresponding to the vacuum wavelength $\lambda$.  A validity condition for this equation is that the Fresnel number $N_{\textrm{F}}$ \cite{SalehTeichBook} obey 
\begin{eqnarray}
N_{\textrm{F}} \equiv \frac{W^2}{\lambda\Delta} \gg 1.
\label{eq:Fresnel_number}
\end{eqnarray}
Here, $W$ is the characteristic transverse length scale for the wave-field being propagated \footnote{Later in the paper, we use the same symbol $W$ to denote the pixel size of the position-sensitive detector that is used to register the intensity of the wave-field.  In general, these length scales---i.e.~the characteristic transverse length scale of the propagating field, and the pixel size of the detector that is used to measure the intensity of the field---will be different from one another.}, and $\Delta\ge 0$ is the distance from (i) the planar exit surface $z=0$ over which the unpropagated wave-field is specified, to (ii) the parallel planar surface $z=\Delta$ over which the intensity of the propagated wave-field is registered using a pixellated position-sensitive detector.  

\begin{figure*}
\includegraphics[trim=0 0 0 0, clip, width=0.9\textwidth]{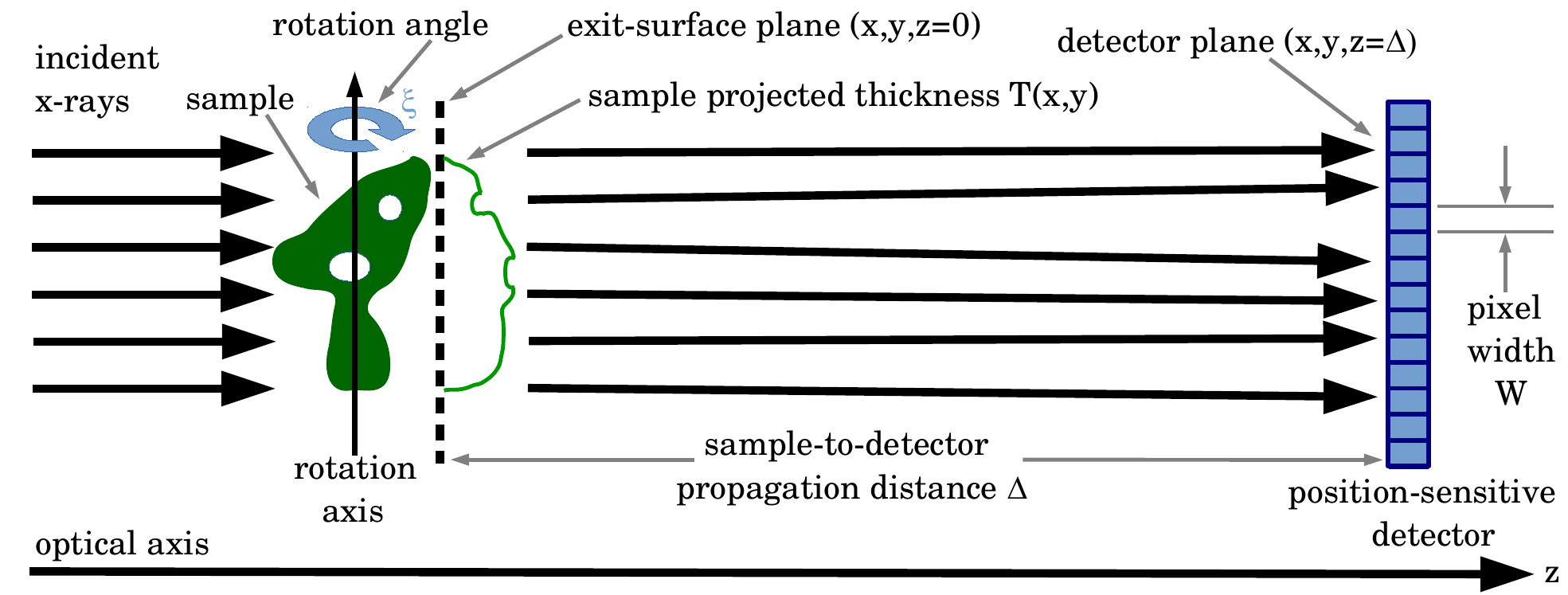}
\caption{A single-material object of projected thickness $T(x,y)$ is illuminated by normally incident $z$-directed monochromatic scalar plane waves of uniform intensity $I_0$, where $(x,y)$ are Cartesian coordinates perpendicular to the optical axis $z$.  The resulting paraxial exit-surface wave-field, over the plane $z=0$, propagates in vacuum through a distance $z=\Delta$.  The associated propagation-based phase contrast image has intensity distribution $I(x,y,z=\Delta)$, sampled using square detector pixels having a width of $W$.  This propagation-based phase contrast image will be sensitive to both the intensity and the phase, of the complex wave-field existing over the exit surface $(x,y,z=0)$ of the sample.  Single-image phase retrieval seeks to recover the projected thickness $T(x,y)$, using the propagation-based phase contrast image as input data.  If a tomographic reconstruction of the sample is performed, the phase-retrieval process can be independently repeated for a number of different angular orientations $\xi$ of the the sample with respect to the indicated rotation axis.  The recovered projected thickness, for this family of angular orientations of the sample, can then be tomographically reconstructed to give a three-dimensional map of the mass density $\rho(x,y,z)$ of the sample. }
\label{fig:BasicDiagram}
\end{figure*}

Following \citet{Paganin2002}, consider a single-material object lying immediately upstream of the plane $z=0$, whose $z$-projection of thickness is given by $T(x,y)$ -- see Fig.~\ref{fig:BasicDiagram}.  The projection approximation \cite{Paganin2006} gives the usual Beer--Lambert law for the intensity $I(x,y,z=0)$ at the exit surface $z=0$ of the object, for the case where the object is illuminated with $z$-directed monochromatic plane waves having uniform intensity $I_0$: 
\begin{eqnarray}
I(x,y,z=0)=I_0 \exp[-\mu T(x,y)].
\label{eq:B}
\end{eqnarray}
\noindent Here, $\mu$ is the linear attenuation coefficient of the single-material  object. The projection approximation also gives an expression for the transverse phase distribution over the exit surface of the object \cite{Paganin2006}:
\begin{eqnarray}
\varphi(x,y,z=0)=-k\delta T(x,y),
\label{eq:C}
\end{eqnarray}
\noindent where $1-\delta$ is the real part of its  complex refractive index
\begin{equation}
n=1-\delta+i\beta    
\end{equation}
and
\begin{equation}
\mu=2k\beta.
\label{eq:Definition_for_mu}
\end{equation}
Note that the single-material object may be generalised to the case of variable mass density $\rho(x,y,z)$, the requirement then being that its complex refractive index have the form
\begin{equation}
    n(x,y,z)=1-A\rho(x,y,z),
\end{equation} 
where $A$ is a fixed complex constant at fixed energy, and ${\textrm{Re}}(A)>0$ \cite{Paganin2004a}. 

Assume vacuum to fill the half space $z\ge 0$ downstream of the object.  Assume the exit-surface wave-field over the plane $z=0$ to propagate through a distance $\Delta > 0$ downstream of the object, with this distance being sufficiently small for the Fresnel number to be much greater than unity.  We may then make the following forward-finite-difference approximation to the longitudinal intensity derivative on the right side of Eq.~(\ref{eq:A}), using the propagation based phase contrast image $I(x,y,z=\Delta)$ of the single-material object in tandem with the estimate for the contact image given by Eq.~(\ref{eq:B}): 
\begin{eqnarray}
\nonumber  \left. \frac {\partial I(x,y,z)}{\partial z} \right|_{z=0} \approx \frac{I(x,y,z=\Delta)-I_0\exp[-\mu T(x,y)]}{\Delta}. \\ 
\label{eq:D}
\end{eqnarray}
If Eqs~(\ref{eq:B}), (\ref{eq:C}) and (\ref{eq:D}) are substituted into Eq.~(\ref{eq:A}), re-arrangement yields the screened Poisson equation \cite{Paganin2002}:
\begin{eqnarray}
\frac{I(x,y,z=\Delta)}{I_0}=\left( 1-\frac{\delta\Delta}{\mu}\nabla_{\perp}^2\right) \exp[-\mu T(x,y)]. 
\label{eq:E}
\end{eqnarray}
The manner in which this has been previously solved is to notice that Fourier transformation turns this partial differential equation into an algebraic equation, via the Fourier derivative theorem.  This leads immediately to the PM \cite{Paganin2002}:
\begin{equation}\label{eq:SPEX}
T(x,y)=-\frac{1}{\mu}\log_e\left(\mathcal{F}^{-1}
\left\{\frac{\mathcal{F}\left[I(x,y,z=\Delta)/I_0\right]}
{1+(\delta\Delta/\mu)(k_x^2+k_y^2)}\right\}
\right).
\end{equation}
\noindent Here $\mathcal{F}$ denotes Fourier transformation with respect to $x$ and $y$ in any convention for which $\nabla_{\perp}$ transforms to $(i k_x,i k_y)$, $\mathcal{F}^{-1}$ is the corresponding inverse Fourier transformation, and $(k_x,k_y)$ are Fourier-space spatial frequencies corresponding to $(x,y)$.  The Fourier-space filter, in the above expression, has the previously-mentioned Lorentzian form. 

When Eq.~(\ref{eq:SPEX}) is directly applied to experimental propagation-based x-ray phase contrast images that are sampled over a Cartesian mesh, and the discrete Fourier transform used to approximate the (continuous) Fourier transform integral, there is an implicit assumption that the object does not contain appreciable spatial frequency information in the vicinity of the Nyquist limit \cite{Press1996} of the mesh.  While this assumption was once typically quite reasonable in most coherent-x-ray-imaging contexts, the exquisitely detailed structures that are now routinely imaged in contemporary x-ray phase-contrast-tomography applications imply that this implicit assumption may now be becoming somewhat less broadly applicable---see e.g.~\citet{Sanchez2012}.  For such applications, and as the following argument will demonstrate,  Eq.~(\ref{eq:SPEX}) overly strongly filters the highest spatial-frequency information that is present in the data.  

With a view to extending the validity of the PM out to the Nyquist limit of the data sampled on a typical pixellated imaging-detector array, recall the following five-point approximation for the transverse Laplacian \cite{Press1996,AbramowitzStegun}, corresponding to a square mesh in which each pixel has a width of $W$:
\begin{eqnarray}
\nonumber W^2\nabla_{\perp}^2 h(x_m,y_n) \approx h(x_{m-1},y_n)+h(x_{m+1},y_n) \\ +h(x_m,y_{n-1})+h(x_m,y_{n+1})-4h(x_m,y_n).
\label{eq:F}
\end{eqnarray}
Here, $h(x,y)$ is a twice-differentiable continuous single-valued function, sampled over a mesh in which each grid element is a square of physical width $W$ metres by $W$ metres.  Hence the mesh locations are given by 
\begin{equation}
(x_m,y_n)=(Wm, Wn),    
\end{equation}
where $m$ and $n$ are integers (mesh indices) that are restricted to the ranges $0\le m\le N_1-1$ and $0\le n \le N_2-1$, with $N_1$ being the number of sample points in the $x$ direction, and $N_2$ being the number of sample points in the $y$ direction.  The key point, here, is that while the fundamental-calculus definition of the transverse derivative considers the mesh step-size $W$ to tend to zero, when working with a discrete grid we are not justified in taking $W$ to be any smaller than the pixel size of the mesh.

With the specified mesh of pixel locations $(x_m,y_n)$ in place, the function $h(x_m,y_n)$ may be expressed in terms of its discrete Fourier transform $H(k_{x,p},k_{y,q})$ \cite{Press1996}: 
\begin{eqnarray}
\nonumber h(x_m,y_n)=\frac{1}{N_1 N_2}\sum_{p=0}^{N_1-1} \sum_{q=0}^{N_2-1} H(k_{x,p},k_{y,q}) \\ \times \exp\left(-\frac{2\pi i m p}{N_1}\right) \exp\left(-\frac{2\pi i n q}{N_2}\right).
\label{eq:G}
\end{eqnarray}
\noindent Here, the discreteness of the sampling grid restricts the allowed spatial frequencies $(k_x,k_y)$ to the Fourier-space mesh
\begin{equation}
(k_{x,p},k_{y,q})=\left(\frac{2\pi p}{N_1 W},\frac{2\pi q}{N_2 W}\right),    
\end{equation}
with $p$ lying in the range $-\tfrac{1}{2}N_1, \cdots, \tfrac{1}{2}N_1$, and $q$ lying in the range $-\tfrac{1}{2}N_2,\cdots,\tfrac{1}{2}N_2$. 

Motivated by the form of the differential operator in Eq.~(\ref{eq:E}), we can show by direct substitution of Eq.~(\ref{eq:G}) into Eq.~(\ref{eq:F}) that (cf.~\citet{Freischlad1986}, \citet{Ghiglia1994} and \citet{arnison2004}):
\begin{eqnarray}
\nonumber (1 &-&  \alpha\nabla_{\perp}^2)h(x_m,y_n) \quad\quad\quad\quad\quad\quad \\ \nonumber &=& \frac{1}{N_1 N_2}\sum_{p=0}^{N_1 - 1} \sum_{q=0}^{N_2-1} H(k_{x,p},k_{y,q}) \\  \nonumber &\times& \left\{1-\frac{2\alpha}{W^2}\left[\cos\left(\frac{2\pi p}{N_1}\right)+\cos\left(\frac{2\pi q}{N_2}\right)-2\right]\right\} \\ &\times& \exp\left(-\frac{2\pi i m p}{N_1}\right)\exp\left(-\frac{2\pi i n q}{N_2}\right),
\label{eq:H}
\end{eqnarray}
\noindent where $\alpha$ is a constant having dimensions of squared length.  Set this constant to the real non-negative number:
\begin{equation}
    \alpha=\frac{\delta\Delta}{\mu}.
    \label{eq:Definition_for_alpha}
\end{equation}
Equation~(\ref{eq:H}) then implies that Eq.~(\ref{eq:E}) may be solved for the projected thickness $T(x,y)$ of the single-material sample, over the lattice of points $(x_m,y_n)$, via the following generalised form of the PM (termed the ``GPM'' henceforth):
\begin{eqnarray}
\nonumber T(x_m,y_n)=-\frac{1}{\mu}\log_e \text{IDFT}^{p\rightarrow m}_{q\rightarrow n}
 \quad \quad \quad \quad \quad \\ \nonumber \times \frac{{\text{DFT}}^{m\rightarrow p}_{n\rightarrow q} [I(x_m,y_n,z=\Delta)/I_0]}{1-\frac{2\alpha}{W^2}\left[\cos\left(Wk_{x,p}\right)+\cos\left(Wk_
{y,q}\right)-2\right]}, \\ \quad \alpha=\frac{\delta\Delta}{\mu},\quad |Wk_{x,p}|,|Wk_{y,q}| \le \pi. 
\label{eq:I}
\end{eqnarray}
\noindent Here, ${\text{DFT}}^{m\rightarrow p}_{n\rightarrow q}$ is the discrete Fourier transform operator, which maps a function $h(x_m,y_n)$ sampled on the real-space lattice $(x_m,y_n)$ to its discrete Fourier transform $H(k_{x,p},k_{y,q})$ sampled on the Fourier-space lattice $(k_{x,p},k_{y,q})$, and  $\text{IDFT}^{p\rightarrow m}_{q\rightarrow n}$ is the corresponding inverse discrete Fourier transform (cf.~Eq.~(\ref{eq:G}); cf.~\citet{Ghiglia1994}, who write a similar expression in the context of the Poisson equation).  Note that operators such as ${\text{DFT}}^{m\rightarrow p}_{n\rightarrow q}$ and the natural logarithm are considered to act from right to left, both in Eq.~(\ref{eq:I}) and for the remainder of the paper, so that e.g.~$\log_e \text{IDFT}^{p\rightarrow m}_{q\rightarrow n} Q$ is equivalent to $\log_e (\text{IDFT}^{p\rightarrow m}_{q\rightarrow n} (Q))$ for any $Q$. Note, also, that the key numerical parameter in Eq.~(\ref{eq:I}) is the dimensionless constant:
\begin{equation}\label{eq:DimensionlessAlphaConstant}
    \frac{\alpha}{W^2}=\frac{\delta/\beta}{4\pi N_{\textrm{F}}}.
\end{equation}

Before proceeding any further, for the sake of clarity we now give a verbal description corresponding to the five-step algorithm expressed in mathematical form by Eq.~(\ref{eq:I}).  This five-step algorithm takes a single propagation-based phase-contrast image $I(x_m,y_n,z=\Delta)$ as input, and yields an estimate for the projected thickness $T(x_m,y_n)$ of a single-material sample that created the measured image, as output.  These five steps are: 
\begin{enumerate}
	\item Normalise the pixellated propagation-based phase-contrast image $I(x_m,y_n,z=\Delta)$ by dividing it through by the uniform intensity $I_0$ of the incident plane-wave illumination.  Note that, if the incident illumination is non-uniform in intensity, this step would correspond to a ``flat field correction'' in which we divide by the position-dependent intensity of the non-uniform illumination.
	\item Take the discrete Fourier transform ${\text{DFT}}^{m\rightarrow p}_{n\rightarrow q}$ of the normalised (flat-field corrected) image, using e.g.~the fast Fourier transform \cite{Press1996}.
	\item Divide by the low-pass Fourier-space filter given by the denominator in the second line of Eq.~(\ref{eq:I}).  Explicitly, this ``GPM Fourier-space filter'' $P_{\textrm{GPM}}$ is:
	\begin{eqnarray}
\nonumber 	  P_{\textrm{GPM}}(k_{x,p},k_{y,q}) \quad\quad\quad\quad\quad\quad\quad\quad\quad\quad\quad\quad\\ =\frac{1}{1-\frac{2\alpha}{W^2}\left[\cos\left(Wk_{x,p}\right)+\cos\left(Wk_
{y,q}\right)-2\right]}.
\label{eq:GPM-filter}
	\end{eqnarray}
	\item Take the inverse discrete Fourier transform $\text{IDFT}^{p\rightarrow m}_{q\rightarrow n}$ of the resulting image.
	\item Take the natural logarithm of the resulting image, then divide by the negative of the linear attenuation coefficient $\mu$.
\end{enumerate}

\begin{figure}
\includegraphics[trim=0 0 0 0, clip, width=0.45\textwidth]{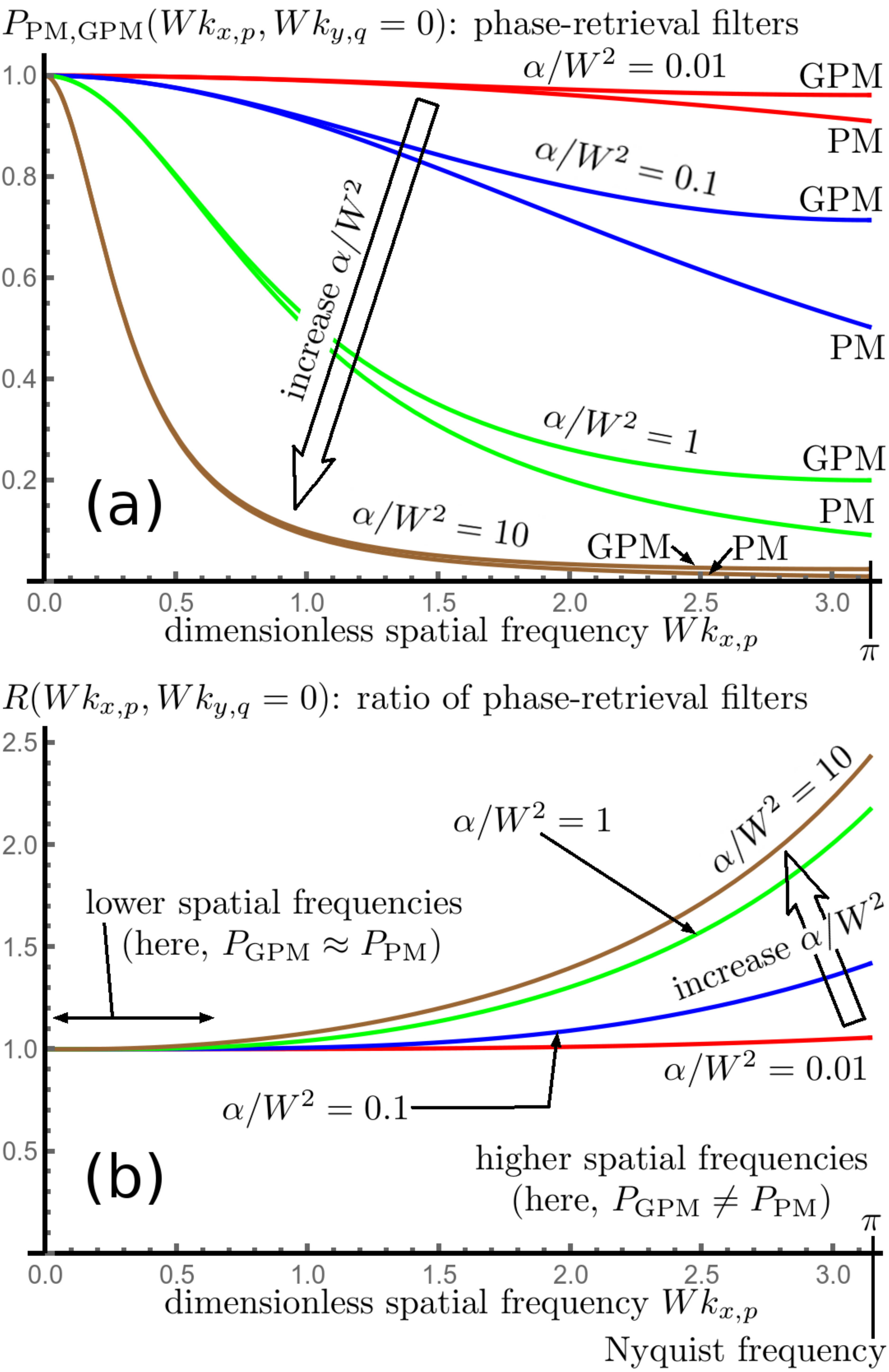}
\caption{(a) One-dimensional cross sections of Fourier-space GPM filter in  Eq.~(\ref{eq:GPM-filter}) and its corresponding PM limiting case in Eq.~(\ref{eq:PM-filter}), for $0 \le W k_{x,p} \le \pi$ and $k_{y,q}=0$, where $W$ is the physical pixel width and $\alpha = \delta \Delta / \mu$. (b) Ratio of filters as given by Eq.~(\ref{eq:ratio-of-filters-in-generic-form}).  
Note that $\alpha/W^2=(4\pi N_{\textrm{F}})^{-1}\delta/\beta$.  Both panels show curves for four different values of the dimensionless parameter $\alpha/W^2$, namely $\alpha/W^2=$0.01 (red), 0.1 (blue), 1 (green) and 10 (brown).}
\label{fig:1D-filters}
\end{figure}

The Fourier-space filter $P_{\textrm{GPM}}(k_{x,p},k_{y,q})$ in Eq.~(\ref{eq:GPM-filter}) is not rotationally-symmetric in the discrete Fourier space of spatial-frequency coordinates $(k_{x,p},k_{y,q})$, but it is nevertheless useful to first plot this filter as a function of only one variable, namely the scaled transverse Fourier-space coordinate $W k_{x,p}$, for the case where $k_{y,q}=0$.  Such plots of $P_{\textrm{GPM}}(W k_{x,p},W k_{y,q}=0)$ are given in Fig.~\ref{fig:1D-filters}(a), for four different cases of the governing dimensionless parameter $\alpha/W^2$ (see Eq.~(\ref{eq:DimensionlessAlphaConstant})).  The dimensionless form of the scaled spatial-frequency coordinate $W k_{x,p}$ here varies from zero to its maximum value (Nyquist frequency) of $\pi$.  Four different values of $\alpha/W^2$ are considered: $\alpha/W^2=0.01$ (red), $\alpha/W^2=0.1$ (blue), $\alpha/W^2=1$ (green) and $\alpha/W^2=10$ (brown).  For each value of $\alpha/W^2$, two curves are provided: the GPM form of the phase-retrieval filter that is given by Eq.~(\ref{eq:GPM-filter}), together with the corresponding PM form of the filter that is given by the rotationally-symmetric expression:
	\begin{eqnarray}
P_{\textrm{PM}}(k_{x,p},k_{y,q}) = \frac{1}{1+\alpha(k_{x,p}^2+k_{y,q}^2)}.
\label{eq:PM-filter}
	\end{eqnarray}
For any given value of $\alpha/W^2$ and for any non-zero spatial frequency,  we see from Fig.~\ref{fig:1D-filters}(a) that the GPM filter is always less strongly suppressing at any specified spatial frequency, in comparison to the corresponding PM filter.  Stated slightly differently, each GPM filter curve always lies above the corresponding PM curve, for any non-zero spatial frequency.  However, it is only at sufficiently-high spatial frequencies that the filters differ appreciably.  As $\alpha/W^2$ increases from the smallest to the largest values in the sequence $\alpha/W^2=0.01,0.1,1,10$ we observe general trends such as: (i) both filters become progressively more strongly filtering, with the GPM filter always being less strongly filtering than the corresponding PM filter; (ii) the discrepancy between the curves is greater at higher spatial frequencies, because the effects of non-zero $W$ become progressively more important for progressively higher spatial frequencies; (iii) the curves converge towards one another, for sufficiently low spatial frequencies, because finite pixel size $W$ is irrelevant for structures that vary so slowly with respect to  transverse position that they are essentially constant over the width $W$ of any single pixel.  Similar general trends may be seen in Fig.~\ref{fig:1D-filters}(b), which plots a one-dimensional cross-section of the Fourier-filter ratio:
	\begin{eqnarray}
R(k_{x,p},k_{y,q})=\frac{P_{\textrm{GPM}}(k_{x,p},k_{y,q})}{P_{\textrm{PM}}(k_{x,p},k_{y,q})}.
\label{eq:ratio-of-filters-in-generic-form}
	\end{eqnarray}

\begin{figure}
\includegraphics[trim=0 30 0 0, clip, width=0.54\textwidth]{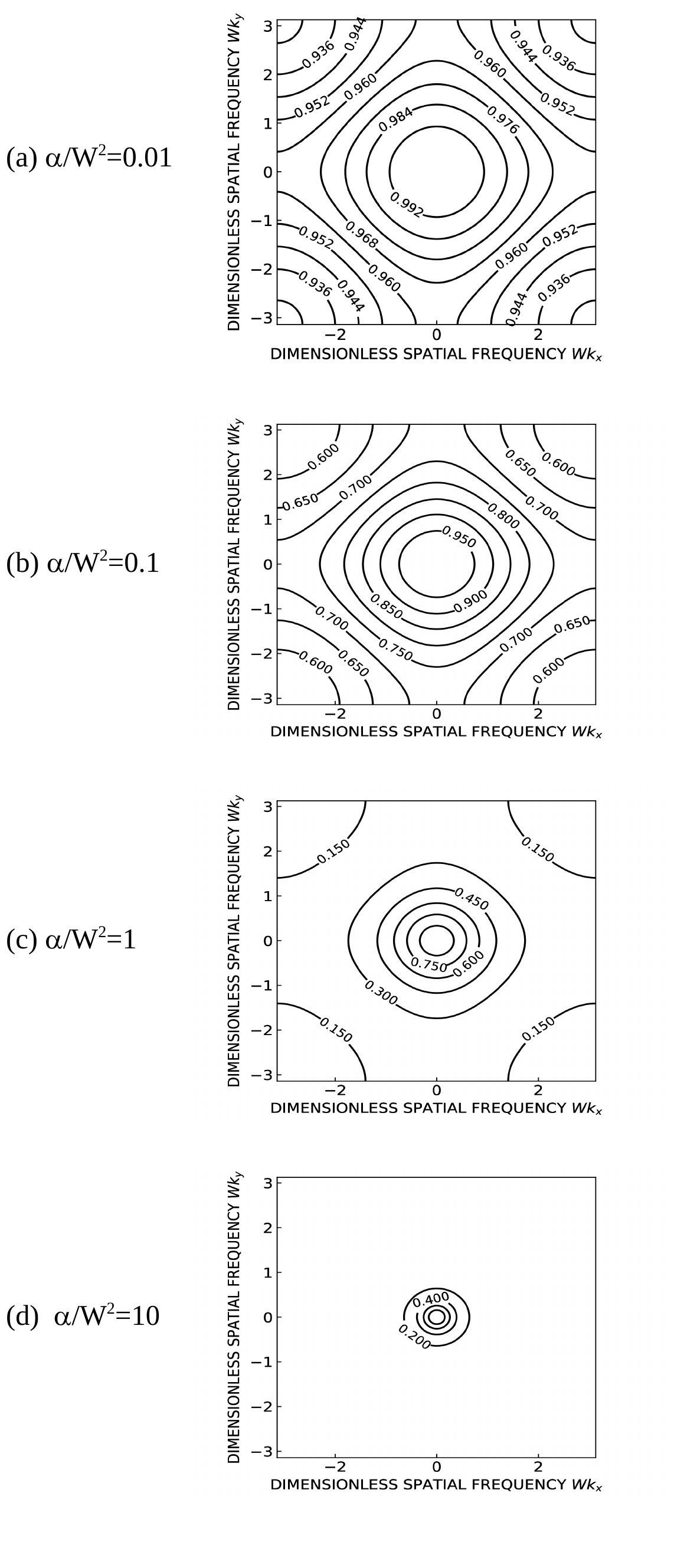}
\caption{Contour plot of GPM Fourier-space filter in  Eq.~(\ref{eq:GPM-filter}), over the full discrete Fourier-space range $-\pi \le W k_x, W k_y \le \pi$, where $W$ is the physical pixel width and $\alpha = \delta \Delta / \mu$. (a) $\alpha / W^2 = 0.01$; (b) $\alpha / W^2 = 0.1$; (c) $\alpha / W^2 = 1$; (d) $\alpha / W^2 = 10$. All plots equal unity at the origin of Fourier space.  Note also that $\alpha/W^2=(4\pi N_{\textrm{F}})^{-1}\delta/\beta.$}
\label{fig:NewFilter}
\end{figure}

As mentioned earlier, the GPM filter is not rotationally symmetric, hence we now consider the fully two-dimensional form of the one-dimensional plots given in Fig.~\ref{fig:1D-filters}(a).  Several two-dimensional plots of the low-pass Fourier-space GPM filter (see Step 3 above, together with Eq.~(\ref{eq:GPM-filter})) are given in Fig.~\ref{fig:NewFilter}, for the same set of four values for the single dimensionless parameter $\alpha/W^2=0.01,0.1,1,10$ considered previously.  Note the transition from (i) near-rotational-symmetry and near-Lorentzian-form close to the origin of Fourier space, to (ii) the symmetry of a square at the edges of Fourier space.  The filter obeys periodic boundary conditions at the edges of the Fourier-space mesh, with each mesh value along the mesh's edges corresponding to a Nyquist frequency 
\cite{Press1996}:
\begin{equation}
W k_{x,p}^{\textrm{max}}, Wk_{y,q}^{\textrm{max}}=\pm\pi.    
\end{equation}
%

In the low-spatial-frequency limit, we may make the second-order Taylor-series approximation
\begin{eqnarray}
\nonumber \cos(Wk_{x,p})\approx 1 - \frac{1}{2}(Wk_{x,p})^2, \\
\cos(Wk_{y,q})\approx 1 - \frac{1}{2}(Wk_{y,q})^2.
\end{eqnarray}
Equation~(\ref{eq:I}) then reduces to the PM with its rotationally-symmetric discrete-Fourier-transform representation of Eq.~(\ref{eq:SPEX}) and the associated Lorentzian filter:
\begin{eqnarray}
\nonumber T(x_m,y_n)\longrightarrow -\frac{1}{\mu}\log_e \text{IDFT}^{p\rightarrow m}_{q\rightarrow n}  \quad \quad \quad \quad \\ \nonumber \times
\frac{{\text{DFT}}^{m\rightarrow p}_{n\rightarrow q} [I(x_m,y_n,z=\Delta)/I_0]}{1+{\alpha\left(k_{x,p}^2+k_{y,q}^2\right)}}, \\ \quad \alpha=\frac{\delta\Delta}{\mu},\quad |Wk_{x,p}|,|Wk_{y,q}| \ll 1.
\label{eq:J}
\end{eqnarray}
\noindent The Lorentzian filter, implied by the above equation, has already been written down in Eq.~(\ref{eq:PM-filter}).  

\section{Computer simulations for propagation-based phase retrieval}
\begin{figure*}
\includegraphics[trim=30 43 20 25, clip, width=1.0\textwidth]{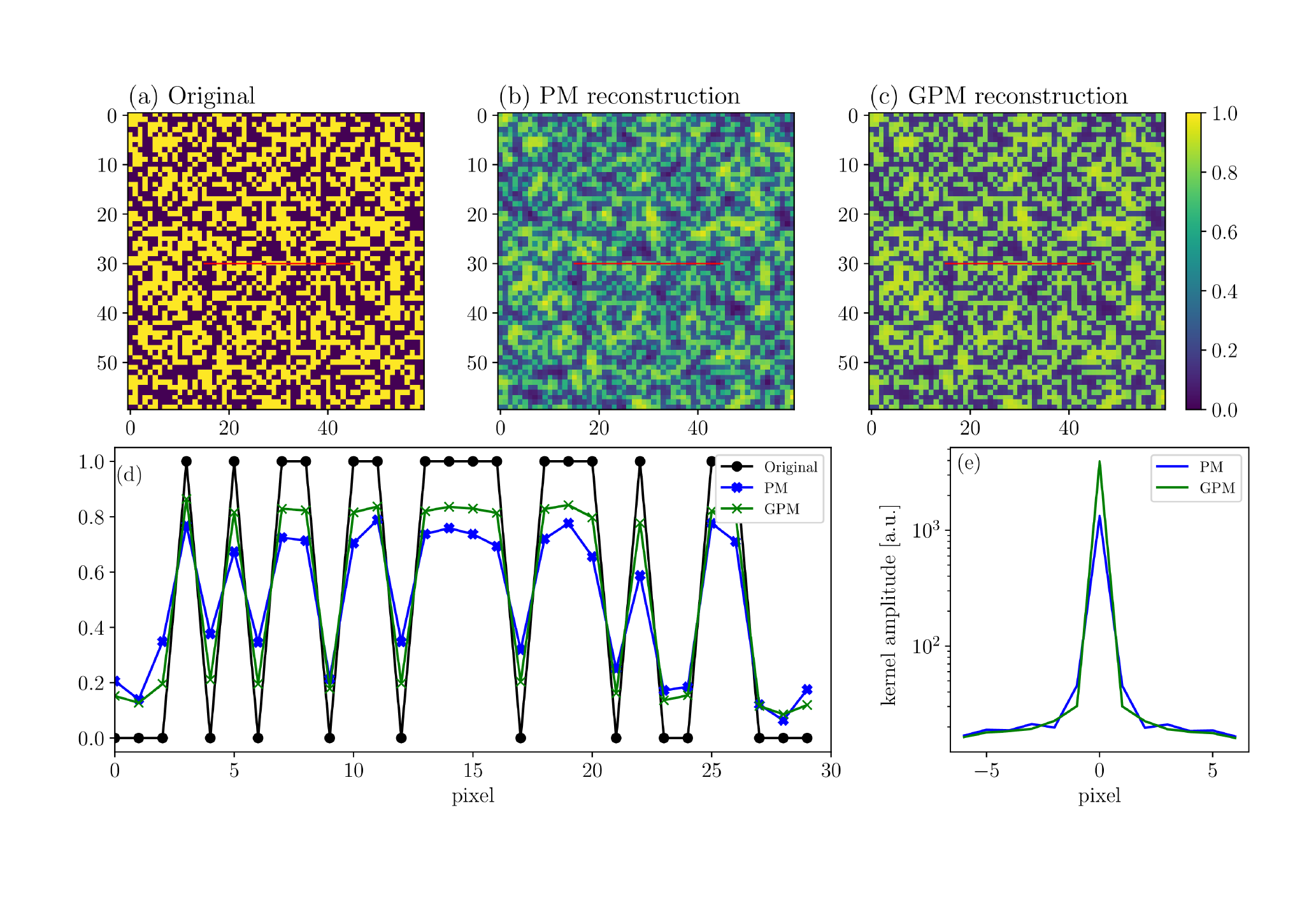}
\caption{Simulated reconstructions using (a) a spatially random binary transmission pattern, before propagation and reconstruction using (b) the standard and (c) generalised approach (see text for details).  Coordinates are given in pixels. (d) Line profile (indicated by a red line in (a-c)) along the original and reconstructed binary patterns. Note that, throughout both this caption and its associated figure, the term ``original'' is a shorthand for ``original image input into the simulation''. Both images and line profiles exhibit a higher contrast for the generalised method.  (e) Line profile of the kernel obtained using a Richardson--Lucy deconvolution \cite{Richardson1972, Lucy1974} between the original array and each of the back-propagated arrays. The generalised method yields a sharper kernel.}
\label{fig:simulation}
\end{figure*}

The efficiency of the generalised phase-retrieval reconstruction method can be asserted by simulating paraxial x-ray propagation through a suitably-high-resolution object, and reconstructing it using both the PM and the GPM. To perform this, we chose a spatially random binary object (Fig.~\ref{fig:simulation}(a)) with an x-ray wavelength of $0.5\ \text{\normalfont\AA}$, $\beta=10^{-9}$, $\delta=5 \times 10^{-7}$, a pixel size of $10\ \mu$m, and a thickness for the object of $40\ \mu$m. 
A simulated unit-amplitude plane wave was transmitted through this object and propagated by a distance of $\Delta=0.1$~m using a near-field propagator. In order to avoid aliasing effects due to the discrete Fourier transform used, the propagation was performed on a $2 \times$ over-sampled object (where the binary random pattern pixels had a $2 \times 2$ size), and the propagated intensity was rebinned (averaging the intensity over $2 \times 2$ pixels) before being back-propagated.  Thus the object-plane image was over-sampled and the corresponding propagated intensity subsequently down-sampled to compensate for the initial oversampling of the object.

Figures~\ref{fig:simulation}(b) and~\ref{fig:simulation}(c) show the intensity of the propagated waves reconstructed using the PM and GPM, respectively, where the obtained thickness maps (normalised to the starting object thickness of $40\ \mu$m) are compared. 
There is a significant improvement in the reconstructed images obtained with the GPM relative to the images reconstructed with the PM. Figure~\ref{fig:simulation}(d) displays line profiles across the images which indicate improvement in the contrast. Since the low-frequency reconstructed images can be approximated as due to a convolution-induced blurring of the original image, we also performed a Richardson--Lucy deconvolution \cite{Richardson1972, Lucy1974} using the original pattern as a reference, which allows us to estimate the point-spread-function kernel relating the reconstructed and original arrays. Figure~\ref{fig:simulation}(e) shows that the GPM yields a sharper kernel.

\section{Experimental results for propagation-based x-ray phase contrast tomography}

Here we give two experimental demonstrations of the ideas developed in the present paper, for 3D propagation-based x-ray phase contrast imaging.  Both experiments were performed at the ID19 microtomography beamline \cite{ID19,ID19-more-recent-paper} of the European Synchrotron (ESRF) in Grenoble, France.

The two samples used for testing the algorithm were (i) a wooden toothpick and (ii) an alumina rod. Images were produced using an incident parallel beam generated by a single-harmonic U17.6 undulator, recorded over $360^{\circ}$ sample rotation on an indirect detector positioned in the near-field condition and consisting of a thin scintillator (i: lutetium aluminium garnet (LuAG) 25 $\mu$m thick; ii: gadolinium gallium garnet (GGG) 100 $\mu$m thick) that was lens-coupled (i: X20 Mitutoyo; ii: X10 Olympus) to a visible-light camera equipped with a $2560 \times 2160$ pixel sCMOS sensor having $6.5\mu\textrm{m}\times6.5\mu\textrm{m}$ pixel size (pco.edge 5.5, PCO, Kelheim, Germany). The principal tomographical and reconstruction parameters, used for both samples, are listed in Table~\ref{LabelForTable}. 

\begin{table}
\begin{tabular}{ |c|c|c|c|c|c|c|c| } 
 \hline
 Sample & $W$~($\mu$m) & $E$~(keV) & $\mathcal{P}$ & $t$~(ms) & $\Delta$ (mm) & $\delta/\beta$ \\ 
 \hline
 Toothpick & 0.32 & 19.6 & 2999 & 30 & 3 & 1961\\ 
 Alumina rod & 0.72 & 35 & 2499 & 35 & 30 & 1797\\ 
 \hline
\end{tabular}
\caption{\label{tab:table-name}Key parameters used for the two propagation-based x-ray phase-contrast tomography experiments: pixel size $W$, energy $E$, number of tomographic projections $\mathcal{P}$, exposure time $t$ for each 2D phase-contrast image, sample-to-detector distance $\Delta$ and delta-to-beta ratio $\delta/\beta$.}
\label{LabelForTable}
\end{table}

\begin{figure*}
 \includegraphics[trim=0 30 0 0, clip, width=1.0\textwidth]{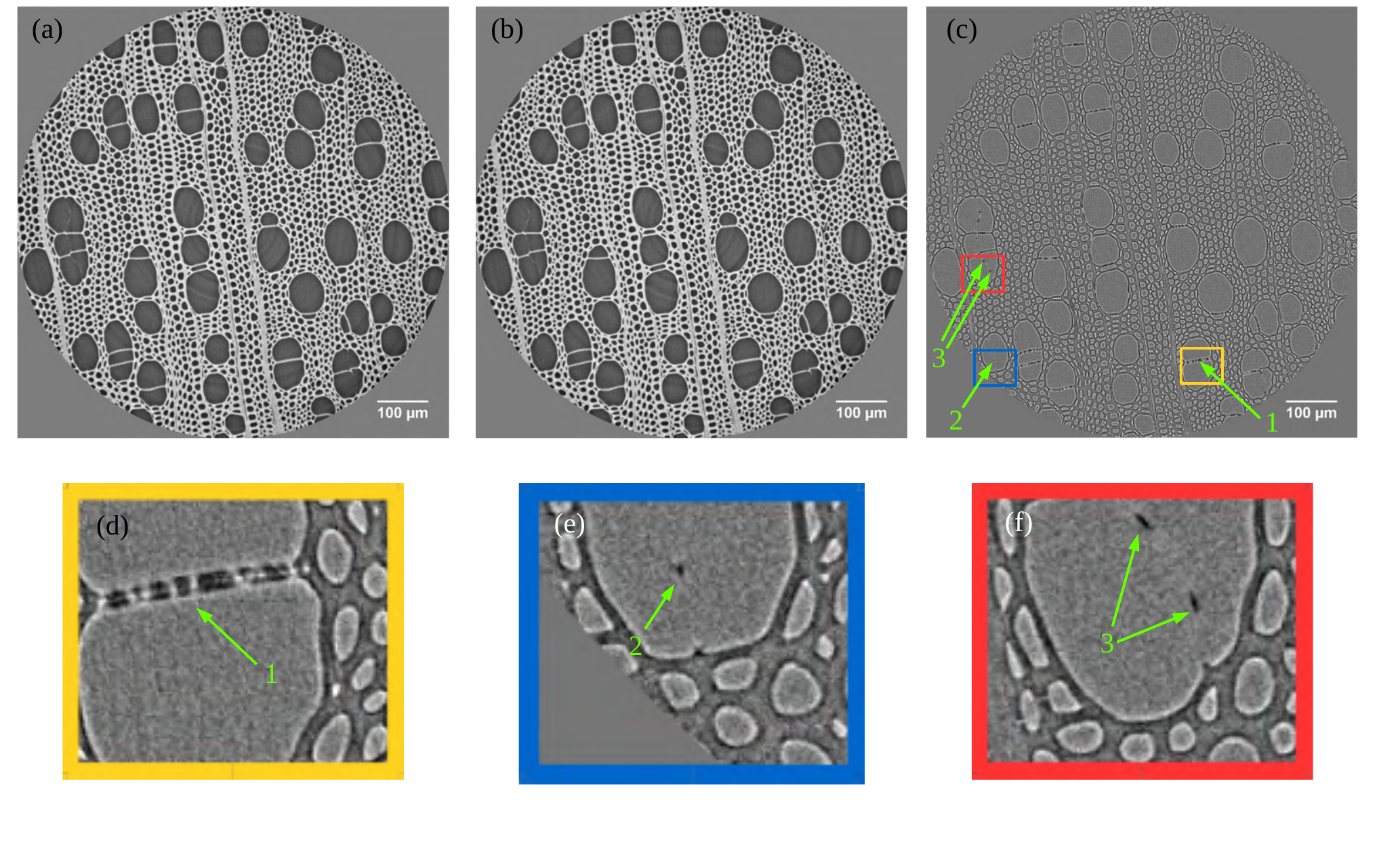}
 \caption{Experimental results for propagation-based x-ray phase contrast tomography, applied to a dry wooden toothpick imaged at 19.6 keV energy. Reconstruction of tomographic slice using (a) PM and (b) GPM, with the pointwise difference between the reconstructions in (a) and (b) being shown in panel (c).  The insets in panel (c), that are bounded by yellow, blue and red rectangles, are magnified in panels (d,e,f), respectively.}
 \label{fig:wood}
  \end{figure*}

The dry wooden toothpick, used for the first experiment, had a nominal diameter of 2 mm.  A $\delta/\beta$ ratio of 1961 was used for the reconstruction, based on the chemical formula for cellulose ($\textrm{C}_6 \textrm{H}_{10} \textrm{O}_5$) at the utilised energy of $E=19.6$ keV. Figure~\ref{fig:wood}(a) shows a reconstruction of one tomographic slice of the sample based on the PM (Eq.~(\ref{eq:J})). Figure~\ref{fig:wood}(b) shows the corresponding GPM reconstruction based on Eq.~(\ref{eq:I}).  These two images look nearly identical to eye, in the greyscale representations of Figs.~\ref{fig:wood}(a) and \ref{fig:wood}(b), however the pointwise difference between these reconstructions, as shown in Fig.~\ref{fig:wood}(c), reveals additional fine-level detail that is present in the GPM reconstruction (Fig.~\ref{fig:wood}(b)) but absent in the PM reconstruction (Fig.~\ref{fig:wood}(a)).  The insets in the difference map (Fig.~\ref{fig:wood}(c)), which are bounded by yellow, blue and red rectangles, are magnified in Figs.~\ref{fig:wood}(d,e,f) respectively.  The additional fine-level detail, which is present in the GPM reconstruction, includes the fine striations in the wood-cell wall that are marked ``1'' in Figs.~\ref{fig:wood}(c,d), the point of increased density within the wood-cell pore that is marked ``2'' in Figs.~\ref{fig:wood}(c,e), and the pair of points with increased density that are labelled ``3'' in Figs.~\ref{fig:wood}(c,f). 

In addition to the difference image in Fig.~\ref{fig:wood}(c) revealing fine spatial detail that is present in the GPM reconstruction but absent from the PM reconstruction, this difference map evidently displays structure related to the Laplacian of the images in Fig.~\ref{fig:wood}(a,b).  We now explain why the obtained  Laplacian-type signal in the difference map (Figs.~\ref{fig:wood}(c,d,e,f)) gives an explicit experimental signature that the GPM indeed gives a higher-spatial-resolution reconstruction than the PM.  Very similar calculations, to those given in the remainder of this paragraph, are presented in e.g.~\citet{Subbarao1995} and \citet{Gureyev2004}, as well as in a number of works related to edge-detection in contexts such as image analysis and machine vision \cite{Edge1,Edge2,Edge3,Edge4,Edge5}.  
Let $f(x,y)$ be a given greyscale image as a function of transverse coordinates $(x,y)$, with
\begin{equation}
f_1(x,y)=f(x,y)\otimes \mathcal{N}_1\exp[-(x^2+y^2)/(2\sigma_1^2)]    
\end{equation}
being the said image blurred by two-dimensional convolution with a Gaussian point-spread function of standard deviation $\sigma_1 > 0$.  Similarly,  
\begin{equation}
f_2(x,y)=f(x,y)\otimes \mathcal{N}_2\exp[-(x^2+y^2)/(2\sigma_2^2)]    
\end{equation}
is the same initial image $f(x,y)$, blurred with a different two-dimensional Gaussian point-spread function of standard deviation $\sigma_2 > \sigma_1$.  Here, $\otimes$ denotes convolution over $x$ and $y$, with
\begin{eqnarray}
\mathcal{N}_{1,2}=\frac{1}{2\pi\sigma_{1,2}^2} 
\end{eqnarray}
being the normalisation constants for the Gaussians, which ensure that each Gaussian integrates to unity.  Note that the condition $\sigma_2 > \sigma_1$ is consistent with the convolution kernels obtained using Richardson--Lucy deconvolution, as shown in Fig.~\ref{fig:simulation}(e), for which the GPM kernel has a standard deviation $\sigma_1$  that is narrower than the standard deviation $\sigma_2$ associated with the PM kernel. The Fourier transforms with respect to $x$ and $y$, of our two blurred images $f_1(x,y)$ and $f_2(x,y)$, are 
\begin{eqnarray}
 \tilde{f}_{1,2}(k_x,k_y) = \tilde{f}(k_x,k_y)\exp[-\tfrac{1}{2}(k_x^2+k_y^2)\sigma_{1,2}^2].    
 \label{eq:HHH}
\end{eqnarray}
Above, note that  (i) we have made use of the convolution theorem of Fourier analysis,  (ii) a tilde denotes Fourier transformation with respect to $x$ and $y$, and (iii) we have used the Fourier-transform convention:
\begin{eqnarray}
  \tilde{f}(k_x,k_y)=\frac{1}{2\pi}\iint f(x,y)\exp[-i(k_x x+k_y y)]\, dx\, dy. \quad\quad
\end{eqnarray}
The exponential term in Eq.~(\ref{eq:HHH}) may be approximated by a second-order Taylor series (inverted parabola)
\begin{equation}
\exp[-\tfrac{1}{2}(k_x^2+k_y^2)\sigma_{1,2}^2]\approx 1-\tfrac{1}{2}\sigma_{1,2}^2(k_x^2+k_y^2),    
\label{eq:JJJ}
\end{equation}
since a sufficiently narrow blur-function in real space will be rather wide in Fourier space, relative to the width of the power spectrum $|\tilde{f}(k_x,k_y)|^2$ of $f$.  If we now substitute Eq.~(\ref{eq:JJJ}) into Eq.~(\ref{eq:HHH}), take the inverse Fourier transform of the resulting expression and then apply the Fourier derivative theorem in reverse, we see that:  
\begin{equation}
    f_{1,2}(x,y)\approx(1+ \tfrac{1}{2} \sigma_{1,2}^2 \nabla_{\perp}^2)f(x,y).
\label{eq:KKK}
\end{equation}
Finally, Eq.~(\ref{eq:KKK}) gives the following expression for the difference between two images of the same function $f(x,y)$, that are obtained at slightly different spatial resolutions:
\begin{equation}
    f_2(x,y)-f_1(x,y)\approx \tfrac{1}{2} (\sigma_2^2-\sigma_1^2) \nabla_{\perp}^2 f(x,y).
\label{eq:LLL}
\end{equation}

\begin{figure}
 \includegraphics[trim=0 0 0 0 0, clip, width=0.44
 \textwidth]{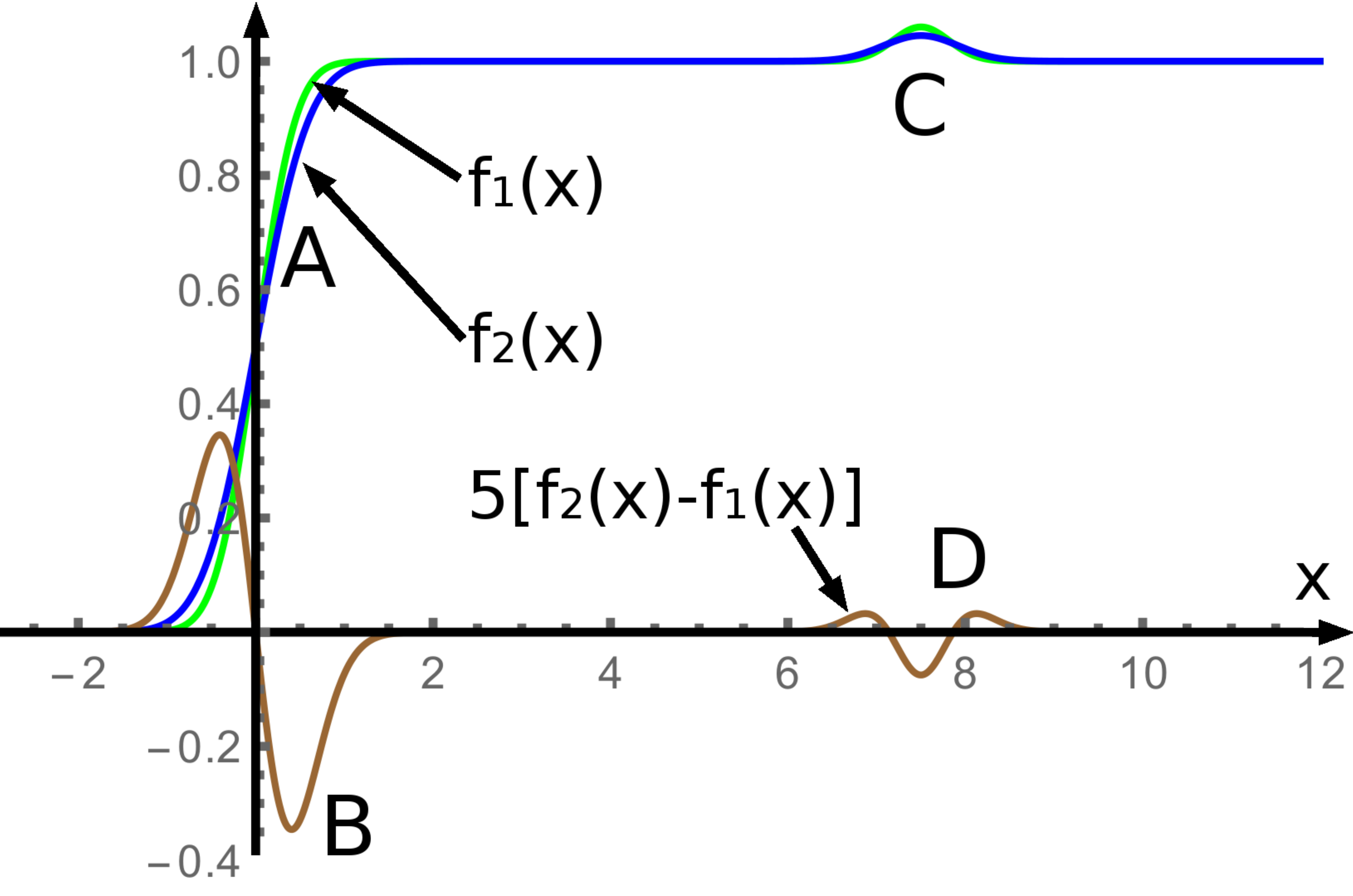}
 \caption{Graphical illustration of Eq.~(\ref{eq:LLL}), showing that the pointwise difference between two images obtained at two slightly different spatial resolutions, is proportional to the Laplacian of either image.  Here, $f_1(x)$ is the image at slightly better spatial resolution (green curve) and $f_2(x)$ is the same image at slightly coarser spatial resolution (blue curve), with both images being considered as functions of one spatial variable $x$ (arbitrary units).  The difference map  $f_2(x)-f_1(x)$ is shown in brown, multiplied by 5 to aid visualisation.}
 \label{fig:signature-from-difference-image}
  \end{figure}

Equation~(\ref{eq:LLL}) shows that the pointwise difference, between two images that are reconstructions of the same function at different spatial resolutions, gives an image that is proportional to the Laplacian of the original image.  This is consistent with the Laplacian signal in Figs.~\ref{fig:wood}(c,d,e,f), but the same result may also be understood conceptually, as sketched in Fig.~\ref{fig:signature-from-difference-image}.  This depicts an image $f(x)$ as a function of one spatial variable $x$, with arbitrary units for $x$, corresponding to a step-like edge $A$ and a ``bump'' $C$.  The green curve for $f_1(x)$ corresponds to slightly better spatial resolution, with the blue curve $f_2(x)$ representing slightly coarser spatial resolution.  When the difference map $f_2(x)-f_1(x)$ is formed, we obtain the brown curve.  The step-like feature $A$ gives a peak-plus-trough feature labelled $B$, which is proportional to the second derivative of the green or blue curves, and which corresponds to the black--white contrast seen at the cell-wall boundaries in Figs.~\ref{fig:wood}(c,d,e,f).  Similarly, the ``bump'' $C$ leads to the feature labelled $D$ in Fig.~\ref{fig:signature-from-difference-image}; such contrast is seen in the dark dots labelled ``2'' and ``3'' in Figs.~\ref{fig:wood}(e,f).     

\begin{figure*}
 \includegraphics[trim=0 0 0 0, clip, width=1.0\textwidth]{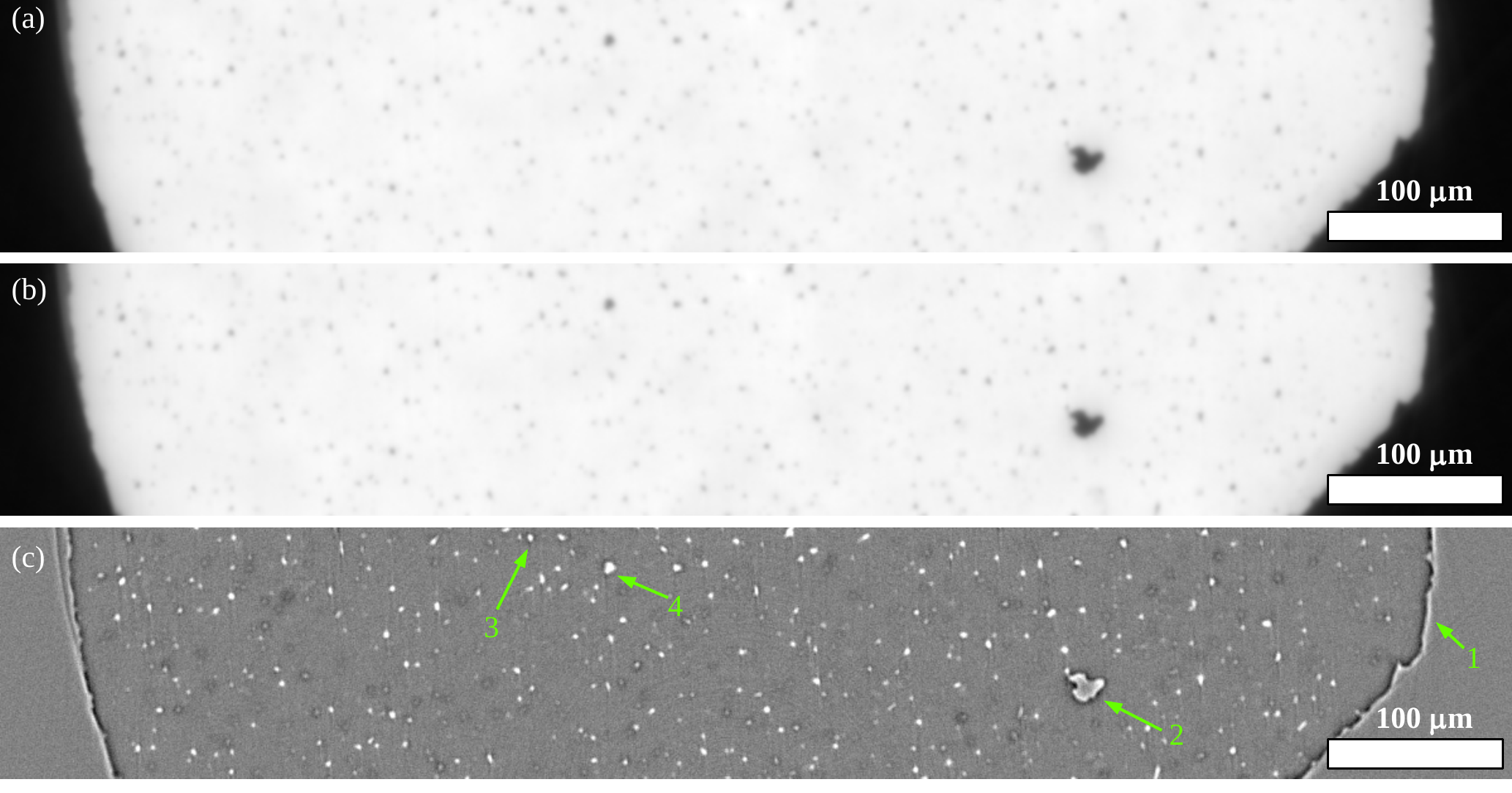}
 \caption{Experimental results for propagation-based x-ray phase contrast tomography, applied to a compacted alumina tube imaged at 35 keV energy. Reconstruction of tomographic slice using (a) PM and (b) GPM, with the difference between the reconstructions in (a) and (b) being shown in panel (c).}
 \label{fig:alumina}
  \end{figure*}

We now turn to the second experiment, which imaged a compacted alumina ($\textrm{Al}_2 \textrm{0}_3$) cylinder, contained within a fused-silica quartz glass capillary having nominal wall thickness of approximately 10 $\mu$m. See Table~\ref{LabelForTable} for other key parameters used in this experiment, leading to the reconstructions shown in Fig.~\ref{fig:alumina}. Figure~\ref{fig:alumina}(a) shows one tomographically reconstructed slice of the alumina cylinder as obtained using the PM, with Fig.~\ref{fig:alumina}(b) giving the corresponding GPM reconstruction.  Figure~\ref{fig:alumina}(c) shows the pointwise difference between the PM and GPM reconstructions.  Alumina pores \cite{Okuma2019} are evident in all reconstructions, with the additional fine spatial detail that is present in panel (b) being rendered visible to the eye in the difference map that is shown in panel (c).   The difference map in Fig.~\ref{fig:alumina}(c) again displays a Laplacian-type signature, consistent with the fact that Fig.~\ref{fig:alumina}(b) has a higher spatial resolution than Fig.~\ref{fig:alumina}(a) (cf.~Eq.~(\ref{eq:LLL}) and Fig.~\ref{fig:signature-from-difference-image}).  As particular examples of the Laplacian-type signature, (i) the black--white edges labelled ``1'' and ``2'' in Fig.~\ref{fig:alumina}(c) correspond to feature $B$ in Fig.~\ref{fig:signature-from-difference-image}, and (ii) the features labelled ``3'' and ``4'' in Fig.~\ref{fig:alumina}(c) correspond to a contrast-reversed form of feature $D$ in Fig.~\ref{fig:signature-from-difference-image}, with the contrast reversal being due to the fact that the centre of such features corresponds to a local trough in sample density rather than a local peak (cf.~feature $C$ in Fig.~\ref{fig:signature-from-difference-image}).  

\section{Discussion}

As stated in the introduction, the work of the present paper was inspired by several experimental investigations \cite{Weitkamp2011,Sanchez2012,mirone2014, Sanchez2013,Irvine2014} that phenomenologically employ unsharp masks or deconvolution to boost high-spatial-frequency information in x-ray phase contrast tomograms whose reconstructions are obtained with the assistance of the PM. This phenomenological modification to the PM often very significantly improves the level of fine-detail clarity in the reconstructions. Is there a fundamental explanation, derivable from an optical-imaging-physics perspective, that casts some light on why the phenomenological high-frequency-boost strategy is so successful?  While we do not claim to give a complete answer to this still-open question, below we argue that the PM-to-GPM transition may go partway to addressing it.  

Consider the ratio $R(k_x,k_y)$ of the GPM and PM Fourier-space filters, as given by Eqs~(\ref{eq:I}) and (\ref{eq:J}).  We have already written this ratio in a generic form, in Eq.~(\ref{eq:ratio-of-filters-in-generic-form}), but in order to now consider it in more detail, we here write it explicitly as:
\begin{eqnarray}
\nonumber R(k_{x},k_{y})=\frac{1+\alpha (k_{x}^2+k_{y}^2)}{1-\frac{2\alpha}{W^2} [\cos(W k_{x})+\cos(W k_{y})-2]}, \\ -\pi/W \le k_{x}, k_{y} \le \pi/W. \quad\quad 
\label{eq:K}
\end{eqnarray}
\noindent This ratio may be viewed as a form of deconvolution mask, here derived from first principles, whose application transforms the PM into the GPM.  As mentioned earlier, such a deconvolution-mask viewpoint may be compared to (and was indeed inspired by) previously-published work which introduced such masks from a phenomenological perspective \cite{Weitkamp2011,Sanchez2012,mirone2014,Irvine2014}. 

\begin{figure}
\includegraphics[trim=0 40 0 0, clip, width=0.52\textwidth]{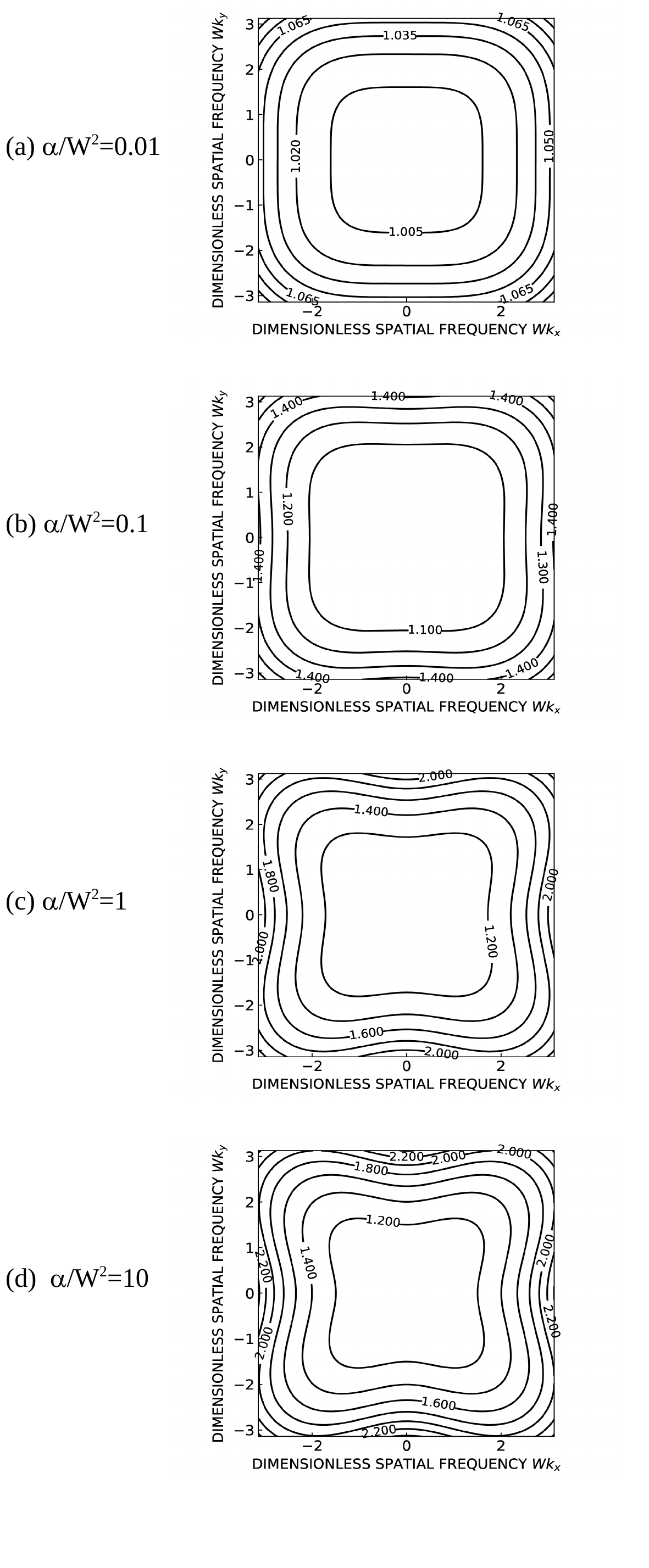}
\caption{Contour plot of the ratio of Fourier-space filters given by Eq.~(\ref{eq:K}). (a) $\alpha / W^2 = 0.01$; (b) $\alpha / W^2 = 0.1$; (c) $\alpha / W^2 = 1$; (d) $\alpha / W^2 = 10$.  All plots equal unity at the origin of Fourier space. Note that $\alpha/W^2=(4\pi N_{\textrm{F}})^{-1}\delta/\beta$.}
\label{fig:Ratios}
\end{figure}

The ratio in Eq.~(\ref{eq:K}) is always greater than or equal to unity, implying that the GPM filter (Eq.~(\ref{eq:I})) suppresses each Fourier component of the measured phase contrast signal, by an amount that is never more than the degree of suppression based on the PM filter (Eq.~(\ref{eq:J})). We have already encountered this point, in the one-dimensional cross-sections that were presented in Fig.~\ref{fig:1D-filters}. To examine this in further detail, see Fig.~\ref{fig:Ratios} for a series of contour plots of the Fourier-filter-ratio in Eq.~(\ref{eq:K}), for the same range of $\alpha/W^2$ values that was used in Figs.~\ref{fig:1D-filters} and \ref{fig:NewFilter}.  The form of these plots---which give a GPM-to-PM filter ratio of unity at the Fourier-space origin, taking values that are progressively greater than unity the further we move away from the Fourier-space origin---gives a partial first-principles justification for the previously cited works boosting high spatial frequencies of tomographic reconstructions based on the PM.  However, we must emphasise that the degree of high-spatial-frequency boost, in the previously cited works \cite{Weitkamp2011,Sanchez2012,mirone2014, Sanchez2013,Irvine2014}, is typically significantly larger than the degree of boost that can be justified using the arguments developed in the present paper.  It is for this reason that we describe our first-principles justification as ``partial'': the GPM gives a reconstruction of fine spatial detail that is superior to that obtained with the PM, but the GPM yields reconstructions that are inferior to those obtained by applying the previously-cited phenomenological approaches \cite{Weitkamp2011,Sanchez2012,mirone2014, Sanchez2013,Irvine2014} to improve the PM by suitably boosting high spatial frequencies.  We conjecture there to be an additional factor or factors that can be used to derive additional high-frequency boosts from first principles, but the nature of these factors remains unanswered by the present investigation.               

In light of the above comments, let us make some additional remarks regarding the  Fourier-filter-ratio plots in Fig.~\ref{fig:Ratios}.  Near the origin of Fourier space, corresponding to coarser spatial detail in the reconstruction, the plots of Fig.~\ref{fig:Ratios} have a plateau of values near unity. Again, this is consistent with the GPM reducing to the PM at sufficiently coarse spatial resolution.  The maximum value $R_{\textrm{max}}$ of the ratio $R(k_x,k_y)$, attained at the corners of the Fourier-space mesh, is given by
\begin{equation}
    R_{\textrm{max}}=\frac{1+2\pi^2\Upsilon}{1+8\Upsilon}, 
\end{equation}
where
\begin{equation}
\Upsilon\equiv\alpha/W^2=(4\pi N_{\textrm{F}})^{-1}\delta/\beta.    
\end{equation}
When $\Upsilon \gg 1$, we see that the maximal Fourier-space boost $R_{\textrm{max}}$ (of the GPM Fourier filter relative to the PM Fourier filter) is by a factor of $\pi^2/4\approx 2.5$: see the corners of Fig.~\ref{fig:Ratios}(d) together with the maximum value taken by the brown curve in Fig.~\ref{fig:1D-filters}(b).  Conversely, when $\Upsilon \ll 1$, then $R_{\textrm{max}}$ tends to unity: see Fig.~\ref{fig:Ratios}(a) together with the red curve in Fig.~\ref{fig:1D-filters}(b).   


If we Taylor expand Eq.~(\ref{eq:K}) to fourth order in spatial frequency, which will be a fair approximation for spatial frequencies that are not too large in magnitude relative to the Nyquist frequency, we obtain
\begin{equation}
R(k_x,k_y) \approx 1+\tfrac{1}{12}\alpha W^2 (k_x^4 + k_y^4).    
\end{equation}
The fact, that the smallest non-constant term in this expansion is quartic in spatial-frequency coordinates, is consistent with the plateau of values close to unity, exhibited by the plots in Fig.~\ref{fig:Ratios} near the Fourier-space origin.  This aligns with the idea that the usual form of the PM works well for coarser spatial detail, but needs the GPM (or some other suitable approach) for better treatment of higher-spatial-frequency detail. The above result also implies that the GPM Fourier filter is approximately equal to $R(k_x,k_y) \approx 1+\frac{1}{12}\alpha W^2 (k_x^4 + k_y^4)$ multiplied by the PM Fourier filter, at least for sufficiently small Fourier frequencies; the corresponding real-space unsharp mask is therefore proportional to the result of applying the non-rotationally-symmetric fourth-order differential operator $\partial^4/\partial x^4 + \partial^4/\partial y^4$ to the PM image, rather than the more usual unsharp mask proportional to the result of applying the rotationally-symmetric second-order differential operator $-(\partial^2/\partial x^2 + \partial^2/\partial y^2)$ to the PM image. 

It is interesting to further examine the conditions under which the GPM differs significantly from the PM.  Take the ratio of Fourier filters in Eq.~(\ref{eq:K}), and evaluate this ratio at the maximum (Nyquist) $x$ and $y$ spatial frequencies
\begin{equation}
k_x^{\textrm{max}}=k_y^{\textrm{max}}=\frac{\pi}{W}.    
\end{equation}
This gives the condition 
\begin{equation}
R\left(k_x=\frac{\pi}{W},k_y=\frac{\pi}{W}\right)=R_{\textrm{max}}=\frac{1+\frac{2\alpha\pi^2}{W^2}}{1+\frac{8\alpha}{W^2}}\geq 1 +\aleph  
\end{equation}
\noindent for the GPM to be significantly different in comparison to the PM.  Here, $\aleph$ is a lower bound on the maximum relative difference, between the ratio of the two filters and unity, which is considered to be ``significant''.  Next we (i) make use of Eqs.~(\ref{eq:Definition_for_mu}) and (\ref{eq:Definition_for_alpha}); (ii) incorporate both the definition of the Fresnel number and its associated requirement as given in Eq.~(\ref{eq:Fresnel_number}).  Hence we obtain the following material-dependent parameter domain for which the GPM is both (i) a significant improvement upon the PM, and (ii) still within the domain of validity for the underpinning transport-of-intensity equation:
\begin{equation}
\frac{\delta}{\beta}\left(\frac{\frac{\pi}{2}-\frac{2}{\pi}}{\aleph}-\frac{2}{\pi}\right)\ge N_{\textrm{F}} \gg 1.
\label{eq:EPMCondition}
\end{equation}
\noindent Thus e.g.~if we choose $\aleph=0.1$, and round numerical factors to the nearest order of magnitude, the above inequalities become the material-dependent conditions:
\begin{equation}
10 \frac{\delta}{\beta} \ge N_{\textrm{F}} \gg 1.
\label{eq:EPMCondition2}
\end{equation}
\noindent Some sample numerical values may be used to illustrate the above expression: if $\delta/\beta=500$, $\lambda=0.5~{\mbox{\normalfont\AA}} =0.5\times 10^{-10}$ m and $W=10~\mu$m$=10^{-5}$ m, then Eq.~(\ref{eq:EPMCondition2}) will be satisfied if $0.4$ mm $\le \Delta \ll 2$ m. If the pixel size is halved to $W=5~\mu$m, leaving all other parameters unchanged, we instead obtain $0.1~{\textrm{mm}} \le \Delta \ll 50$ cm.  If the condition in Eq.~(\ref{eq:EPMCondition}) is violated then more sophisticated methods than either the PM or the GPM will need to be employed, including but not limited to (i) holotomography \cite{Cloetens1999}, (ii) approaches based on the first Born and Rytov approximations \cite{Gureyev--RytovBorn}, (iii) approaches based on the contrast-transfer-function formalism \cite{Gureyev2004b--LinearAlgorithmsPaper,Turner2004}, and (iv) the variety of approaches that are both reported upon and compared in \citet{Yu2018}.   

One further point should be made regarding unsharp masks and deconvolution. The GPM will still benefit from additional unsharp masking or deconvolution, to further boost high-spatial-frequency detail, since---among other reasons beyond the scope of the present investigation, such as truncation of the effects of Fresnel diffraction to ignore the presence of multiple Fresnel-diffraction fringing---the GPM does not explicitly take source-size-induced blurring into account.  The degree of sharpening required for GPM-reconstructed images will necessarily be less pronounced than that which has been needed for PM-reconstructed images.  In this context we point out that the image-blurring effect due to finite source size may be modelled by making the following replacement in Eq.~(\ref{eq:E}) \cite{beltran2018}:
\begin{equation}
\frac{\delta}{\mu}\longrightarrow \frac{\delta}{\mu}-\frac{2 S^2}{\Delta}.
\label{eq:Beltran2018}
\end{equation}
Here, $S$ is the radius of the effective incoherent point-spread function at the detector plane, that is due to source-size blurring (cf.~\citet{Gureyev2004}).  The above replacement transforms Eq.~(\ref{eq:E}) into a Fokker--Planck form \cite{Risken1989,MorganPaganin2019,PaganinMorgan2019} for which $2 S^2/\Delta$ plays the role of an effective  diffusion coefficient \footnote{See, in particular, the special case of Eq.~(59) in the paper by \citet{PaganinMorgan2019}, in which their ``effective diffusion coefficient'' $D_{\textrm{eff}}(x,y)$ is considered to be independent of transverse coordinates $(x,y)$.  The resulting Fokker--Planck equation is mathematically identical in form to the screened Poisson equation given when Eq.~(\ref{eq:E}) of the main text is modified by the replacement that is indicated in Eq.~(\ref{eq:Beltran2018}).}.  This simple algebraic replacement may be carried through all of the calculations of the present paper, thereby incorporating a partial source-size deconvolution into the analysis.

It is also interesting to observe that we can introduce a real parameter $\tau$, which lies between zero and unity inclusive, that continuously deforms the PM algorithm in Eq.~(\ref{eq:J}) ($\tau = 0$), into the GPM algorithm in Eq.~(\ref{eq:I}) ($\tau = 1$), via:
\begin{eqnarray}
\nonumber T(x_m,y_n;\tau)=-\frac{1}{\mu}\log_e \text{IDFT}^{p\rightarrow m}_{q\rightarrow n} \quad \quad \quad \quad \quad \\  \times \frac{{\text{DFT}}^{m\rightarrow p}_{n\rightarrow q} [I(x_m,y_n,z=\Delta)/I_0]}{1+{\alpha\left(k_{x,p}^2+k_{y,q}^2\right) - \frac{2\alpha\tau}{W^2} \Phi(k_{x,p},k_{y,q})}}, 
\label{eq:TUNABLE}
\end{eqnarray}
\noindent where $0\le\tau\le 1$ and
\begin{align}
\nonumber
\Phi(k_{x,p},k_{y,q})\equiv\cos(Wk_{x,p})+\cos(Wk_{y,q})-2 \\ +\tfrac{1}{2}(Wk_{x,p})^2+\tfrac{1}{2}(Wk_{y,q})^2.    
\end{align}

We close this discussion by returning, once again, to the theme of unsharp-mask image enhancement.  As pointed out earlier, the work of the present paper was inspired by the success of applying unsharp-mask processing and related concepts, to improve the fine spatial detail apparent in PM-based reconstructions.  In this context, it is perhaps fitting to ``come full circle'' by pointing out that the difference maps, such as those shown in Figs.~\ref{fig:wood}(c,d,e,f) and \ref{fig:alumina}(c), may themselves be viewed as a form of unsharp mask.  We have already argued and observed that such difference maps emphasise the fine spatial details that are present in the GPM reconstructions, but which are suppressed in the corresponding PM-based reconstructions.  The reason that this additional fine detail is not evident in comparing the GPM to the PM reconstructions, is that this additional fine detail is of sufficiently small amplitude to be difficult to discern by eye in greyscale representations such as those shown in Figs.~\ref{fig:wood}(b) and \ref{fig:alumina}(b). This last-mentioned observation arises from the physiological limitations of the human eye, which can only discriminate on the order of 30 different greyscale levels \cite{Kreit2013}.  It is also worth pointing out, in the present phase-retrieval imaging context, that it is often of utility to decompose physical models of samples---such as spatial distributions of complex refractive index---into a sum of (i) a spatially slowly-varying function that may and in general will have a large magnitude, and (ii) a spatially rapidly-varying function that has a small amplitude relative to the amplitude of the slowly-varying component.  Such decompositions are used in e.g.~\citet{Gureyev--RytovBorn, Gureyev2004b--LinearAlgorithmsPaper} and \citet{Nesterets2008}, together with references therein.  Bearing all of the above points in mind, let $T_{\textrm{PM}}(x,y)$ denote a PM-based reconstruction of a projected thickness (or density) distribution, with $T_{\textrm{GPM}}(x,y)$ denoting the corresponding GPM reconstruction.  We then write the identity:
\begin{eqnarray}
T_{\textrm{GPM}}(x,y)=T_{\textrm{PM}}(x,y)+[T_{\textrm{GPM}}(x,y)-T_{\textrm{PM}}(x,y)]. \quad\quad \label{eq:GPM-minus-PM-identity}
\end{eqnarray}
The term in square brackets, on the right-hand side of Eq.~(\ref{eq:GPM-minus-PM-identity}), is the ``difference map'' plotted in Figs.~\ref{fig:wood}(c,d,e,f) and \ref{fig:alumina}(c).  Considering this to be a form of unsharp mask that is induced by the transition from the PM to the GPM, we may follow the formulation given in e.g.~\citet{AdrianSheppard2004} to modify Eq.~(\ref{eq:GPM-minus-PM-identity}) as follows: 
\begin{eqnarray}
T(x,y;s)\equiv T_{\textrm{PM}}(x,y)+s[T_{\textrm{GPM}}(x,y)-T_{\textrm{PM}}(x,y)]. \quad\quad \label{GPM-induced-unsharp-mask}
\end{eqnarray}
Here, $s$ is a real parameter.  The case $s=0$ corresponds to using the PM, with $s=1$ corresponding to the GPM.  When $s > 1$ we may consider $T(x,y;s > 1)$ to be an unsharp-mask image-sharpened representation of $T(x,y)$.  In a similar vein, we may also consider Eq.~(\ref{eq:TUNABLE}), in which $\tau$ is now taken to be a real parameter that is greater than unity, to constitute a form of unsharp-mask image enhancement that is directly based upon the GPM.

\section{Some avenues for future work}

Since \citet{Beltran2010,Beltran2011} reported signal-to-noise ratio (SNR) boosts of up to 200 in utilising the PM in a tomographic setting, relative to absorption contrast, there has been some interest in the ``SNR boosting'' properties of the PM \cite{Nesterets2014,Gureyev2014,Kitchen2017,Gureyev2017}. Of particular note is the result that the SNR boost has 0.3 $\delta/\beta$ as an approximate upper bound under the assumption of Poisson statistics \cite{Nesterets2014,Gureyev2014}, with the SNR boost being even more favourable for very low sample-exposure times \cite{Kitchen2017,Clark2019}.  Since dose is proportional to the square of SNR, dose reductions of $300^2 =90,000$ or more are in principle possible with the PM \cite{Kitchen2017}.  This implies that tomographic analyses are possible using much less dose (or, equivalently, much lower acquisition times) than previously required for a single two-dimensional projection.  This reduced dose is of importance in the context of medical imaging, while in an industrial-inspection product-quality-control context (or security screening context) it enables significant increases in throughput speed due to the associated increase in source effective-brilliance; cf.~e.g.~the recent achievement of over 200 x-ray phase-contrast tomograms per second \cite{200tps}, incorporating PM-based data processing.  In light of the above comments, it would be interesting to see how the previously published analyses for SNR boost and associated dose reduction are altered by passage from the PM to the GPM.  It appears likely that SNR boosts will be reduced somewhat if the GPM is used in place of the PM, in accord with the trade-off between noise and spatial resolution \cite{GureyevNRU,GureyevNRU2020}. 

Another interesting avenue for future work begins with the previously mentioned observation that, from version 2.1 onwards, the ANKAphase \cite{Weitkamp2011} implementation of the PM incorporates a deconvolution mask to boost fine spatial detail.  This deconvolution filter $R_{\textrm{ANKA}}(k_x,k_y)$ takes the Fourier-space form \footnote{See \texttt{https://imagej.nih.gov/ij/plugins/ankaphase/ ankaphase-userguide.html} for an updated form of the ANKAphase software, which extends the form reported in \citet{Weitkamp2011}, to incorporate the optional image-restoration deconvolution filter that is reproduced in Eq.~(\ref{eq:ANKAphase_filter}) of the main text.}
\begin{equation}\label{eq:ANKAphase_filter}
R_{\textrm{ANKA}}(k_x,k_y)=\frac{1+c}{c+\exp[-\pi\sigma^2(k_x^2+k_y^2)]},    
\end{equation}
where $c$ is a dimensionless constant and $\sigma$ is a characteristic width.  In light of the findings of the present paper, it would be interesting to consider replacing the ANKAphase deconvolution filter with
\begin{equation}\label{eq:ANKAphase_filter_revised}
\tilde{R}_{\textrm{ANKA}}(k_x,k_y)=\frac{1+c}{c+\exp[-\sigma^4(k_x^4+k_y^4)]},    
\end{equation}
since the fourth-order Taylor expansion of the deconvolution filter would then agree with the fourth-order Taylor expansion of $R(k_x,k_y)$, provided that
\begin{equation}
\frac{12\sigma^4}{1+c}=W^2\alpha.   
\end{equation}

It would also be interesting to investigate any utility that the present work may have, in the context of tomographic image segmentation \cite{HermanBook,OhserSchladitzBook}. For example, would a tomographic segmentation---of difference maps such as those shown in Figs.~\ref{fig:wood}(c) or \ref{fig:alumina}(c)---reveal additional fine morphological detail compared to tomographic segmentation of unsharp-mask image-sharpened PM-based reconstructions?  What advantage (if any) arises for image segmentations that are based on the GPM-induced unsharp-mask image-enhancement given by the $s > 1$ case of Eq.~(\ref{GPM-induced-unsharp-mask}), or the $\tau > 1$ case of Eq.~(\ref{eq:TUNABLE})?  What advantage (if any) arises for image segmentations that are based on taking existing unsharp-mask image-enhancement strategies, and applying them to GPM-based rather than PM-based reconstructions?  The efficacy of the various approaches to segmentation, as listed above, could be quantitatively compared via their application to segmentation-derived quantities such as porosity, surface density, Euler-number density, mean-curvature density, mean coordination number, fibre-direction distributions etc.~\cite{OhserSchladitzBook}.

One more avenue for future research would be to replace Eq.~(\ref{eq:F}) with a higher-order discrete approximation to the transverse Laplacian.  For example, Eq.~(25.3.30) of \citet{AbramowitzStegun} makes it clear that, while we have a Taylor-series truncation error on the order of $W^2$ in the five-point approximation in Eq.~(\ref{eq:F}), their nine-point approximation in Eq.~(25.3.31) \cite{AbramowitzStegun} gives a better estimate whose truncation error is on the order of $W^4$.  Use of such higher-order approximations may  lead to improved forms of Eq.~(\ref{eq:I}).

We end these indications of possible avenues for future work, by noting that both (i) the two-image TIE-based phase-retrieval method of \citet{Paganin1998} (see also Sec.~4.5.2 of \citet{Paganin2006} as well as Sec.~4 of \citet{Zuo2014}), which does not make the assumption of a single-material object, and (ii) the single-image differential-phase-contrast version of the PM \cite{Paganin2004b}, which does make such an assumption, would benefit from an analogous treatment of differential operators under the discrete Fourier transform, to that used in the passage from Eqs.~(\ref{eq:J}) to (\ref{eq:I}).  Perhaps the former method (namely that which is based on Ref.~\cite{Paganin1998}) may become of increasing utility in an x-ray imaging setting, given both recent advances in semi-transparent detectors, and the fact that this method does not need the single-material assumption upon which the PM relies. 

\section{Conclusion}

A simple extension was given for the method of \citet{Paganin2002} (PM), for phase--amplitude reconstruction of single-material samples using a single paraxial x-ray propagation-based phase contrast image obtained under the conditions of small object-to-detector propagation distance. This improves the reconstructed contrast of very fine sample features, using an approximation that is derived from a first-principles perspective.  This investigation was motivated by, and partially explains from a fundamental perspective, the success of several papers incorporating unsharp masking or related techniques to boost fine spatial detail in reconstructions obtained using the PM \cite{Weitkamp2011,Sanchez2012,mirone2014, Sanchez2013,Irvine2014}. 
\section*{Acknowledgements}
Financial support by the Experiment Division of the ESRF for DMP to visit in 2017 and 2018 is gratefully acknowledged.  DMP acknowledges fruitful discussions with Mario Beltran, Carsten Detlefs, Timur Gureyev, Marcus Kitchen, Kieran Larkin, Thomas Leatham, Kaye Morgan, Tim Petersen, James Pollock, Manuel S\'{a}nchez del R\'{i}o and Paul Tafforeau. DMP dedicates this paper to the memory of Wendy Micallef-Paganin.


\bibliography{GPM2020}

\end{document}